# DESCRIPTION OF QUANTUM BEAMS USING A STATIONARY INCOHERENT SUPERPOSITION OF WAVE PACKETS

Xavier Oriols and Jordi Suñé
Departament d'Enginyeria Electrònica (ETSE)
Universitat Autònoma de Barcelona
08193- Bellaterra SPAIN

**Abstract**

An original entity is defined to study quantum beams of conserved particles. The beam is described as a Stationary Incoherent Superposition of Wave packets (SISOW). Mean values are time-independent confirming that the SISOW describes a stationary flux of particles. The SISOW is a general entity that contains the Hamiltonian eigenstates as a particular limit. The decomposition of the SISOW into wave packets provides intrinsic dynamic information which is not accessible from Hamiltonian eigenstates. The SISOW provides and adequate framework to study mesoscopic systems.

Among the various theoretical approaches used to study open systems, the one proposed by Landauer and Büttiker (LB) [1-2] has become a standard to treat coherent transport in mesoscopic systems and it has had a great success in many experiments [3]. The Hamiltonian eigenstates are the main components of this approach since the conductance of the system is assumed to be proportional to the transmission coefficient of these states. Moreover, the open system conditions can easily be modelled through the asymptotic boundary conditions of the Hamiltonian eigenstates. On the other hand, the study of quantum systems in terms of wave packets has had little practical use for a long time because the experimental analysis at the relevant time scale was not possible. However, new advances in the physics and chemistry of laser interaction with atoms have brought wave packets into the limelight [4]. In this framework, the theoretical study of the dynamics of wave packets is regarded with renewed interest and many examples of wave packet phenomenology can be found in the recent literature [5-9]. In this letter, we present an original description for a beam of particles using a constant flux of time dependent wave packets. By means of an incoherent superposition of time-dependent wave packets, we construct a new quantum entity with properties that are in between those of stationary Hamiltonian eigenstates and those of the individual wave packets. With the Hamiltonian eigenstates, it shares stationary properties such as a uniform time-independent current and a stationary probability presence profile. On the other hand, the wave packet decomposition provides a detailed dynamical information not accessible from Hamiltonian eigenstates.

We will focus upon the description of open systems that interchange locally conserved particles with their environment. We are interested in describing the stationary flux of particles impinging upon a localised potential, keeping at the same time the possibility of extracting consistent information about the dynamics of the interaction. In order to present our proposal in a simple scenario, we will study a one-dimensional system located between two particle reservoirs, considering the injection from the left reservoir (the emitter). For a time independent potential profile, a time dependent wave packet, $\Psi_{t_o}(x,t)$, can be written as a linear superposition of Hamiltonian eigenfunctions $\varphi_E(x)=<x|E>$ as:

$$\Psi_{t_o}(x,t) = \langle x | \Psi_{t_o}(t) \rangle = \int_0^{+\infty} dE \cdot a(E) \cdot \varphi_E(x) \cdot e^{-i\frac{E \cdot (t-t_o)}{\hbar}} \tag{1}$$

a(E) being a complex quantity . By using the superposition for the numerical calculation of $\Psi_{t_o}(x,t)$, the open-system boundary conditions are directly inherited from the Hamiltonian eigenstates that compound the wave packet and eventual spurious reflections at the limits of the system are avoided. From an intuitive point of view, a beam of particles can be thought as a constant flux of identical time-dependent wave packets each one leaving the emitter reservoir and entering into the system at different times $t_o$. Following this idea, we propose a description based on an incoherent superposition of identical wave packets that enter into the system distributed in time with uniform probability. For such an incoherent superposition, we follow a formulation based on the density matrix, which in the energy representation is given by:

$$\rho_\Psi(E,E') = \frac{1}{2\pi\hbar} \int_{-\infty}^{+\infty} dt_o \cdot \langle E | \Psi_{t_o}(t) \rangle \langle \Psi_{t_o}(t) | E' \rangle = \frac{1}{2\pi\hbar} \int_{-\infty}^{+\infty} dt_o \cdot a(E) \cdot a^*(E') \cdot e^{-i\frac{(E-E')(t-t_o)}{\hbar}} \tag{2}$$

The constant $1/2\pi\hbar$ comes from normalization in the sense that the trace is equal to unity. Although such a normalization is not actually required for an open system, it gives the correct results in the limit of Hamiltonian eigenstates (i.e. $a(E) = \delta(E - E')$). Expression (2) is the definition of a quantum entity that we will call a SISOW (Stationary Incoherent Superposition Of Wave packets). Noting that $\int dt_o \cdot \exp(i(E'-E)t_o/\hbar) = 2\pi\hbar \cdot \delta(E'-E)$, it immediately follows that:

$$\rho_\Psi(E,E') = |a(E)|^2 \delta(E'-E) \tag{3}$$

Thus, even though the density matrix associated to time-dependent wave packets is not diagonal, the density matrix of a SISOW is indeed diagonal. On the other hand, the mean value of any observable $\hat{A}$ is given by:

$$\langle \hat{A} \rangle = tr(\hat{\rho} \cdot \hat{A}) = \int_0^\infty dE \cdot |a(E)|^2 \cdot A(E,E) \tag{4}$$

where A(E,E) are the diagonal elements of $\hat{A}$ in the energy representation. In particular, we notice that $\langle \hat{A} \rangle$ does not depend on time and this confirm that the SISOW is an adequate entity to describe stationary situations. Moreover, in the limit of wave packets infinitely narrow in energy, we recover the stationary results corresponding to the eigenstates.

Let us now work out an example to further highlighting the stationary properties of the SISOW. The probability presence profile associated to the SISOW, $Q_\psi(x)$, can be computed by using expression (4), or alternatively, as an incoherent sum over the probability presence of the whole ensemble of wave packets. Thus:

$$Q_\psi(x) = \frac{1}{2\pi\hbar} \int_{-\infty}^{\infty} |\Psi_{to}(x,t)|^2 \cdot dt_o = \int_{-\infty}^{\infty} |a(E)|^2 \cdot |\varphi_E(x)|^2 \cdot dE \tag{5}$$

This expression can also be recognised as the dwell time of the particle at each position, and it reveals the relation between the local density of states and the dwell time which was previously discussed by Iannaccone [10]. Continuing with the description of stationary properties, maybe the most important one for a stationary beam of particles is the current. In this regard, the current density associated to a SISOW, $J_\Psi$, can be written as:

$$J_\Psi = \frac{1}{2\pi\hbar} \int_{-\infty}^{\infty} J_{t_o}(x,t) \cdot dt_o = \frac{1}{2\pi\hbar} \int_{-\infty}^{\infty} |a(E)|^2 \cdot T(E) \cdot dE = \frac{T_\Psi}{2\pi\hbar} \qquad (6)$$

where $J_{t_o}(x,t)$ is the current density associated to the $t_o$-wave packet, and $T(E)$ is the transmission coefficient of the eigenstate $\varphi_E(x)$. The current density of a SISOW is time-independent, uniform, and proportional to the transmission coefficient of the wave packet, $T_\Psi$. In figure 1, we have represented the probability presence and current densities of a beam of electrons impinging upon a 20 Å thick 0.04 eV height square potential barrier. Gaussian wave packets with a central energy of 0.015 eV and spatial standard deviation of 70 Å have been used to build up the SISOW. The charge and the current have been numerically computed according to equations (5) and (6) respectively. The results corresponding to the SISOW are compared with those of the eigenstate associated to the central energy. In both cases, the current is uniform and time independent, and the charge profile is stationary. However, in spite of these similarities, the results show qualitative differences. For example, the oscillations of the eigenstate charge in the far emitter region disappear when a SISOW is considered. In general, we can say that the SISOW is a more flexible entity that contains the energy eigenstate as a monochromatic limit. In other words, an eigenstate can be considered as a SISOW with $a(E) = \delta(E - E')$ (i.e. with an infinite spatial dispersion).

Let us now consider the dynamic properties of the SISOW. We distinguish between two levels of dynamic information: (i) the *average dynamic information* obtained as a result of equation (4), and (ii) the *intrinsic dynamic information* obtained by decomposition of the SISOW into the ensemble of identical time-dependent wave packets. Let us begin with the analysis of the linear momentum distribution. First of all, for each single wave packet the momentum distribution is given by $\left|\Psi_{t_o}(p,t)\right|^2$, with $\Psi_{t_o}(p,t) = \langle p | \Psi_{t_o}(t) \rangle$. This distribution is time dependent: it remains on a region of positive momenta while the wave packet is incident, and shifts towards negative momenta as soon as the wave packet is partially reflected. Moreover, this distribution shows very interesting classically forbidden effects as recently noticed by Brouard and Muga when the scattering of a wave packet upon a barrier is considered [9]. On the contrary, as it also happens for the stationary eigenstates, the momentum distribution associated to the SISOW, $P_\psi(p)$, is time independent:

$$P_\psi(p) = \frac{1}{2\pi\hbar} \int_{-\infty}^{\infty} \left|\Psi_{t_o}(p,t)\right|^2 \cdot dt_o = \int_{-\infty}^{\infty} |a(E)|^2 \cdot |\boldsymbol{j}_E(p)|^2 \cdot dE \qquad (7)$$

where $\boldsymbol{j}_E(p) = \langle p | \boldsymbol{j}_E \rangle$. However, although $P_\psi(p)$ does not depend on time, equation (7) explicitly shows that it is obtained as an average of the wave packet momentum distribution. This decomposition in terms of $\left|\Psi_{t_o}(p,t)\right|^2$ allows the consideration of intrinsic dynamic information of the beam that is not accessible from an eigenstate. In order to deepen into this intrinsic dynamic information of a SISOW, let us make use of the method of quantum trajectories proposed by de Broglie and Bohm [11]. In addition to reproducing the standard quantum mechanical measurable results, the Bohm trajectories provide a valuable intuitive insight of the underlying physical processes. Within the Broglie-Bohm formalism, the velocity at each time and position, $v_{to}(x,t)$ is computed as:

$$v_{to}(x,t) = J_{to}(x,t)/|\Psi_{to}(x,t)|^2 \qquad (8)$$

Let us now discuss the dynamic properties of a SISOW by evaluating the velocity. First, the Bohm velocity of an eigenstate, $v_{j_E}(x) = T(E)/2p\hbar |j_E(x)|^2$, is always positive and time independent. Analogously, it can be straightforwardly shown from equation (5) and (6) that the average velocity of a SISOW is $v_\Psi(x) = T_\Psi/2p\hbar Q_\Psi(x)$. In both cases, there is a unique velocity for each position which does not represent the actual velocity of the quantum particles of the beam at any time because it averages the velocities corresponding to incident and reflected particles. On the contrary, the velocity of a particular wave packet at each position depends on time, $v_{to}(x,t)$ and is basically positive when the wave packet is incident and negative when it has already been reflected by the barrier. The main advantage of the SISOW is that it allows computing the average velocity $v_\psi(x)$ as an average over the velocities $v_o(x,t)$ of the ensemble of wave packets. In this regard, we can obtain a probability distribution of velocities given by:

$$P_\Psi(v,x) = \frac{1}{2p\hbar} \int_{-\infty}^{\infty} |\Psi_{to}(x,t)|^2 \cdot d(v - v_{to}(x,t)) \cdot dt_o \qquad (9)$$

Let us point out several properties of this distribution. $P_\Psi(v,x)$ is always positive and time independent by construction. The probability presence density, $Q_\psi(x)$, and the current density, $J_\Psi$, are easily obtained from $P_\Psi(v,x)$. As an example, in figure 2a), we have depicted the position-velocity distribution, $P_\psi(x,v)$, of the SISOW described in figure 1. The oscillations of this distribution just before the barrier are directly related with the oscillations of the probability presence in figure 1. Moreover, the broadening of the velocity distribution at these positions can be understood as a direct consequence of the current uniformity. Since the probability presence decreases in the minima of the pre-barrier oscillations, the velocity distribution has to be broader to preserve the current uniformity. This broadening of the velocity distribution of the SISOW in front of the barrier can be directly related to the transient collisional enhancement of the high momentum components of the individual wave packets recently reported by Brouard and Muga [11]. Finally, we want to notice the presence of the tunnelling ridge that accounts for the partial transmission through the barrier. In figure 2b) we have represented the velocity of the eigenstate with an energy equal to the central energy of the SISOW. In this case, there is only one possible velocity at each position, so that this velocity can only represent an average between incident and reflected particles.

At this point, we want to make a brief discussion of the potential advantages of the application of the SISOW to the modelling of transport phenomena. In particular, we want to point out the possible benefits of a generalised LB approach in terms of SISOWs instead of Hamiltonian eigenstates. The LB approach [1,2] has become the standard to deal with coherent transport in mesoscopic systems. Following the guidelines of this approach, a coherent system with injection from several energy channels can also be studied by considering an ensemble of single-particle SISOWs with different central energies and invoking the concept of thermal reservoirs to incorporate them into a density matrix that describes the whole system. Notice that a SISOW can model beams of particles either wide or narrow in energy (i.e. to model the "size of the particle" in the sense discussed by Fischetti [12]) and we can approach the monochromatic limit as much as we want without loosing the time-dependent information of the wave packet that forms the SISOW. The capability of this generalised LB approach to deal with beams of particles with femptosecond resolution can be of great interest in the present research fields of current fluctuations and transient phenomena in mesoscopic systems. We point out that our approach directly takes into account the discreteness of the charge that evolves inside the system (i.e. the individual wave packets). It can be easily shown that expression (4) is time-dependent (i.e. noisy) if the injection of wave packets inside the system is considered at discrete intervals of time. Moreover, another advantage of our proposal is that it can be extended to study non-coherent transport within a simple and versatile framework. In this regard, it has been repeatedly pointed out in the literature that the treatment of dissipative processes in far from equilibrium conditions requires non-diagonal elements in the density matrix [13]. Since the density matrix of a single wave packet is non-diagonal, the presence of non-diagonal elements in the density matrix of the global system can be 'intuitively' described as a process of creation or annihilation (due to different relaxation processes) of the wave packets that compound the SISOWs (notice that creation or annihilation of Hamiltonian eigenstates do not provide non-diagonal elements). Several approaches can be found in the literature to rigorously determine the population of these non-diagonal elements of the density matrix. Among them, we mention the one proposed by Fausto Rossi and co-workers [14] where energy-relaxation and dissipative processes are introduced via microscopic models by means of a generalised semiconductor Bloch equation in a quite simple framework. From a practical point of view, the viability of the present approach to study quantum transport in mesoscopic systems was already demonstrated in a previous work where a quantum Monte Carlo simulator based on wave packets and Bohm trajectories was developed to deal with tunnelling transport in heterostructures devices [15]:

The conductance characteristics of resonant tunnelling diodes has been obtained by self-consistently solving this generalised LB approach with the Poisson equation. That work was the origin of our present interest in transport in terms of wave packets. The present letter also serves as a formal justification of that work [15].

In conclusion, we have presented an original description for the transport of conserved particles in terms of a constant flux of time dependent wave packets. By means of an incoherent superposition of time-dependent wave packets, we construct a new quantum entity (the SISOW) with properties that are in between those of stationary Hamiltonian eigenstates (uniform time-independent current and a stationary probability presence profile), and those of the dynamic wave packets. It has been shown that the mean value of any observable in a system described by a SISOW is time-independent. This confirms that the SISOW is an adequate entity to describe stationary fluxes of particles. The SISOW is a general entity that contains the Hamiltonian eigenstate in the monochromatic limit. The main advantage of a SISOW, however, is that it is possible to decompose it into time-dependent individual wave packets. This decomposition provides information about intrinsic dynamic properties of the beam that are not accessible when the system is modelled by Hamiltonian eigenstates. This dynamic information can be very useful to study the phenomenology of a beam of particles in the nowadays experimentally accessible femtosecond regime. In particular, a generalisation of the LB approach in terms of SISOWs instead of Hamiltonian eigenstates has been discussed. Transient, dissipative or noise phenomena in mesoscopic systems has been pointed out as possible interesting applications of this approach.

**Acknowledgement**

This work has been partially supported by the DGES under project number PB97-0182.

**References**

[1] R.Landauer, Philos. Mag. 21, 863 (1970)
[2] M. Büttiker, Y.Irmy, R.Landauer, and S.Pinhas, Phys. Rev. B, 31, 6207 (1985)
[3] Supriyo Datta, *Electronic transport in mesoscopic systems*, Cambridge University press, (1995)
[4] B.M.Garraway and K.A.Suominent, Rep. Prog. Phys., 58, 365 (1995)
[5] T.C.Weinacht *et al.* Phys. Rev. Lett., 80, 5508 (1998)
[6] I.Sh. Averbukh, Phys. Rev. Lett., 77, 3518 (1996)
[7] V.G.Lyssenko *et al.* Phys. Rev. Lett., 79, 301 (1997)
[8] D.R.Bitouk and M.V.Fedorov, Phys. Rev. A, 58, 1195 (1998)
[9] S.Brouard and J.G.Muga, Phys. Rev. Lett., 81, 2621 (1998)
[10] G.Iannaccone, Phys. Rev. B, 51, 4727 (1995)
[11] P.R.Holland, *The Quantum Theory of motion* (Cambridge University press, Cambridge 1993); D.Bohm and B.J.Hilley, *The undivided Universe* (Toutledge, New York 1993); X.Oriols et al., Phys. Rev A, 54, 2594 (1996)
[12] M.V. Fischetti, J. Appl. Phys., 83, 270 (1998)
[13] F.Rossi, A.di Carlo and Paolo Lugli, Phys. Rev. Lett. , 80, 3348 (1998)
[14] W.R.Frensley, Rev. of Mod. Phys., 62, 745 (1990)
[15] X.Oriols *et al.* Appl. Phys. Lett., 72, 806 (1998)

**Figure caption**

**Fig 1:** Beam of electrons with central energy 0.015 eV impinging upon a rectangular barrier of 20 Å width and 0.04 eV height. The probability presence and current densities of a beam of particles are represented. The system is modelled by an eigenstate (dashed lines) and by a SISOW (solid lines). Vertical lines represent the barrier.

**Fig 2:** Dynamic properties of the system described in figure 1: (a) De Broglie-Bohm velocity distribution at each position for the SISOW in a contour plot and (b) velocity at each position for an eigenstate. Vertical lines represent the barrier .The velocity distribution of a SISOW can disjoin incident and reflected particles, however, the velocity of the eigensate is an average of the velocities of the incident and reflected particles.

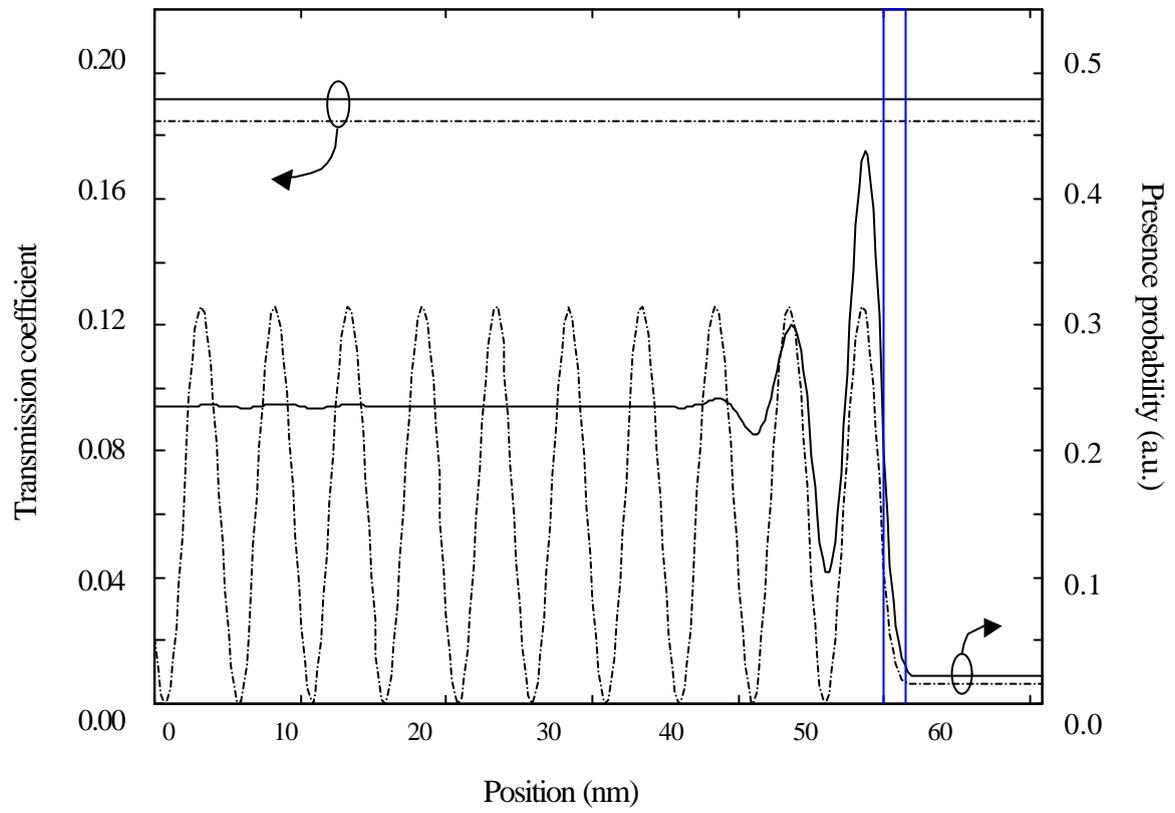

**Figure 1:**  Physical Review Letters                                        X.Oriols et al.

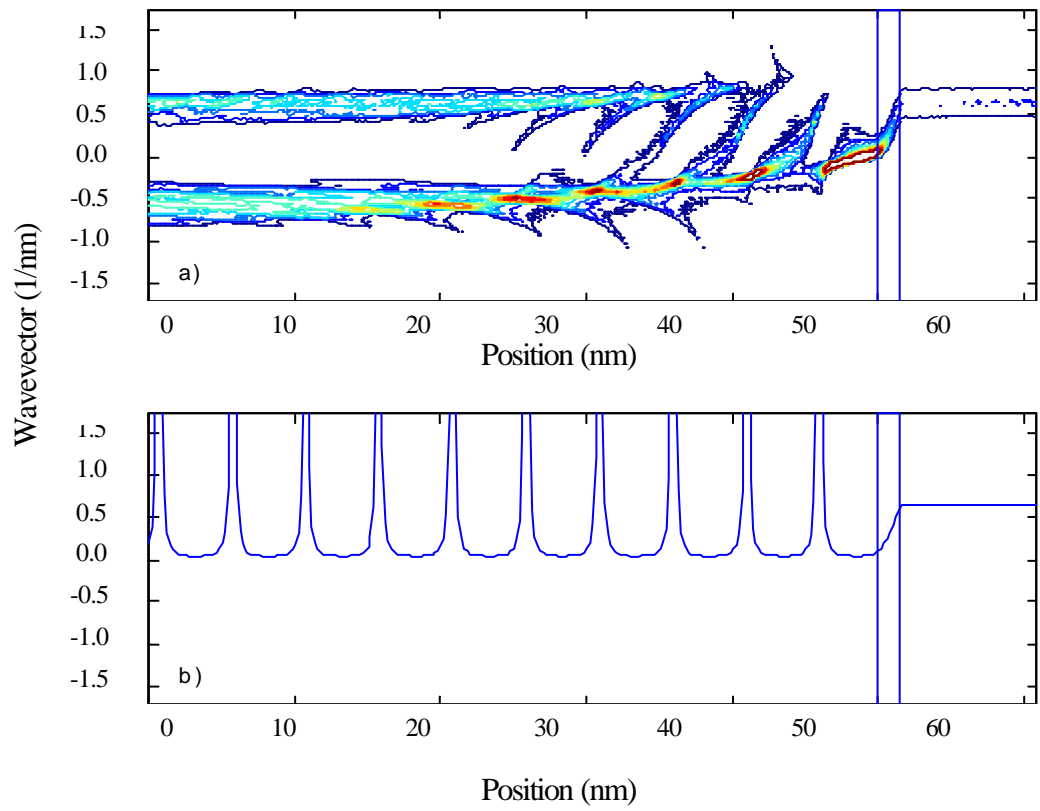

**Figure 2:** Physical Review Letters  X.Oriols et al.